\documentclass[12pt]{JHEP3}

\usepackage{cite}
\usepackage{amssymb}
\usepackage{amsbsy}
\usepackage{graphicx}

\newcommand{\be}{\begin{equation}} \newcommand{\ee}{\end{equation}}
\newcommand{\bea}{\begin{eqnarray}} \newcommand{\eea}{\end{eqnarray}}
\newcommand{\el}{\nonumber \\}
\newcommand{\re}[1]{(\ref{#1})}

\newcommand{\adot}{\dot{a}}
\newcommand{\addot}{\ddot{a}}
\newcommand{\bdot}{\dot{b}}
\newcommand{\bddot}{\ddot{b}}

\newcommand{\brt}[1]{[#1]}
\newcommand{\rmd}{\textrm{d}}
\newcommand{\ack}{\acknowledgments}

\renewcommand{\H}{H_a}
\newcommand{\HH}{H_a^2}
\newcommand{\Hb}{H_b}
\newcommand{\HHb}{H_b^2}

\newcommand{\Hd}{\frac{\addot}{a}}
\newcommand{\Hbd}{\frac{\bddot}{b}}

\newcommand{\PRD}[1]{{\it Phys. Rev.} {\bf D #1}}
\newcommand{\PRL}[1]{{\it Phys. Rev. Lett.} {\bf #1}}

\newcommand{\NPB}[1]{{\it Nucl. Phys.} {\bf B #1}}
\newcommand{\PLB}[1]{{\it Phys. Lett.} {\bf B #1}}
\newcommand{\MNRAS}[1]{{\it Mon. Not. Roy. Astron. Soc.} {\bf #1}}
\newcommand{\APJ}[1]{{\it Astrophys. J.} {\bf #1}}
\newcommand{\APJS}[1]{{\it Astrophys. J. Suppl.} {\bf #1}}
\newcommand{\JHEP}[1]{{\it JHEP} {\bf #1}}
\newcommand{\JCAP}[1]{{\it JCAP} {\bf #1}}
\newcommand{\CQG}[1]{{\it Class. Quant. Grav.} {\bf #1}}

\renewcommand{\AA}[1]{{\it Astron. Astrophys.} {\bf #1}}

\title{Dark energy and decompactification\\in string gas cosmology}

\author{Francesc Ferrer \\ CERCA, Department of Physics, Case Western Reserve University, \\ 10900 Euclid Avenue, Cleveland, OH 44106-7079, USA \\
Email: \email{francesc.ferrer@case.edu}}
\author{Syksy R\"{a}s\"{a}nen \\
Rudolf Peierls Centre for Theoretical Physics, University of Oxford,
\\ 1 Keble Road, Oxford, OX1 3NP, UK \\
Email: \email{syksy.rasanen@iki.fi}}

\abstract{We study the stability of extra dimensions
in string gas cosmology at late times.
Vacuum energy and, interestingly, baryons
lead to decompactification after they become dynamically
important. The string gas can stabilise the
effect of baryons, but not that of vacuum energy.
However, we find that the interplay of baryons and strings
can lead to acceleration in the visible dimensions,
without the need for vacuum energy.}

\keywords{Cosmology of Theories beyond the SM}

\preprint{hep-th/0509225}

\begin{document}

\section{Introduction}

String gas cosmology (SGC)
\cite{Kripfganz:1988, Brandenberger:1989, Tseytlin:1991, Sakellariadou:1995, Alexander:2000, Easson:2001, Easther:2002mi, Watson:2002, Easther:2002, Bassett:2003, Brandenberger:2003, Watson:2003, Easther:2003, Watson:2003uw, Patil:2004, Watson:2004vs, Kaya:2004, Berndsen:2004, Easther:2004, Danos:2004, Brandenberger:2005a, Biswas:2005a, Patil:2005a, Patil:2005b, Easson:2005, Berndsen:2005, Battefeld:2005, Kanno:2005, Brandenberger:2005b}
(see \cite{Brandenberger:2005b} for an introductory overview and
a more comprehensive list of references)
is a cosmological scenario motivated by string theory.
In SGC, unlike in most applications of string theory, all
spatial dimensions are treated on an equal footing: they are all
compactified and start out small, and filled with a hot
gas of branes of all allowed dimensionalities.
Also in contrast to most higher-dimensional proposals,
SGC aims to explain not only why some dimensions are hidden,
but also why the number of visible dimensions is three
(see \cite{Durrer:2005, Karch:2005} for other proposals
along the same lines).

In the simplest versions of SGC there are nine spatial
dimensions compactified on tori, all with initial sizes
near the self-dual radius $\sqrt{\alpha'}\equiv l_s$.
The branes can wind around the tori. The energy of the winding modes
increases with expansion due to the tension of the branes,
and this resists expansion. As the universe expands and cools
down, winding and anti-winding modes annihilate, allowing further
expansion. A simple counting argument suggests that $p$-branes
and their anti-branes cannot find each other to annihilate in more
than $2p+1$ spatial dimensions, so at most $2p+1$ dimensions
can become large. For $p=1$, corresponding to strings, this is three
spatial dimensions.
(Some quantitative studies of brane gases have cast doubt on this
qualitative argument, see \cite{Sakellariadou:1995, Easther:2002, Bassett:2003, Easther:2003, Easther:2004, Danos:2004} for different analyses.)

It has been shown \cite{Patil:2004, Patil:2005a} that strings
winding around the extra dimensions can stabilise them at the self-dual
radius in the radiation dominated early universe while the
visible dimensions expand, even when the dilaton is stabilised
(see also \cite{Kripfganz:1988, Watson:2003, Berndsen:2004, Brandenberger:2005a, Patil:2005b, Berndsen:2005, Kanno:2005}). However, it was noted that the
extra dimensions decompactify\footnote{We use the word
'decompactify' in the common, technically incorrect, sense of
becoming macroscopically large. No change of topology is implied,
and all spatial dimensions remain compact at all times.}
if the universe is dominated by four-dimensional dark matter
(a problem discussed early on in \cite{Kripfganz:1988, Weiss:1986}).
It was concluded that ordinary dark matter has to be replaced by
extra-dimensional dark matter (strings winding around the
extra dimensions) to obtain viable late-time cosmology.

We take a closer look at late-time cosmology and decompactification.
We note that even if dark matter is extra-dimensional, ordinary
baryons will destabilise the extra dimensions after they
become dynamically important. However, we find
that the strings which stabilise the extra dimensions
during the radiation dominated era can counter the
destabilising push of four-dimensional pressureless
matter, whether it is baryons or dark matter. The competition between
the push of dust and the pull of strings cannot make the
extra dimensions static, but it can lead to oscillations
around the self-dual radius with decreasing amplitude, so
that the extra dimensions can be regarded as effectively
stabilised. Like dust, vacuum energy will lead to decompactification,
but the strings cannot help there. However, we discover a
possibility for obtaining acceleration without any extra dark
energy component: the oscillations induced by dust and strings
can involve transitions between deceleration and acceleration,
even when the energy density of the universe is dominated by matter.

In section 2 we present the calculation of destabilisation by
baryons and stabilisation by strings. We then show how the vacuum
energy destabilises the extra dimensions and how the competition
between four-dimensional matter and strings can lead to
acceleration. In section 3 we discuss the relation to
observations and summarise our results.

\section{Dark energy and decompactification}

\subsection{Set-up}

\paragraph{The metric and the equation of motion.}

We will consider a ten-dimensional spacetime, with all
nine spatial directions compactified on tori\footnote{The
important part about the topology is that it must have
one-cycles for strings to wind around. For
discussion of more complex compactifications, see
\cite{Easson:2001, Easther:2002mi}.}. Six of the 
dimensions remain small, while three are large.
We take the metric to be the simplest generalisation of the
spatially flat Friedmann-Robertson-Walker universe, homogeneous
and separately isotropic in the visible and the extra dimensions:
\bea \label{metric}
  \rmd s^2 = - \rmd t^2 + a(t)^2 \sum_{i=1}^3 \rmd x^i\rmd x^i + b(t)^2 \sum_{j=1}^6 \rmd x^j\rmd x^j \ ,
\eea

\noindent where $i=1\ldots3$ labels the visible dimensions
and $j=1\ldots6$ labels the extra dimensions. Our convention
is that for the small dimensions the value $b=1$ corresponds to
extra dimensions at the self-dual radius $l_s$.
For the three large dimensions, we make the more
convenient choice of $a=1$ corresponding to the
Big Bang Nucleosynthesis (BBN) era, specifically, to
the time when $\rho_m/\rho_{\gamma}=10^{-6}$ 
(and $\rho_{\gamma}\sim$(MeV)$^4$).

We  are interested in late-time behaviour, so we assume
that the dilaton has been stabilised in a way that
leaves the equation of motion of the metric unconstrained,
so that it reduces to the Einstein equation
(see \cite{Berndsen:2004, Brandenberger:2005a, Patil:2005b, Berndsen:2005, Kanno:2005})
\bea \label{eom}
  G_{\mu\nu} = \kappa^2 T_{\mu\nu} \ ,
\eea

\noindent where $G_{\mu\nu}$ is the Einstein tensor,
$\kappa^2$ is the 10-dimensional gravitational
coupling and $T_{\mu\nu}$ is the energy-momentum tensor
(we lump the cosmological constant together with vacuum
energy as part of the energy-momentum tensor).

Given the symmetries of the metric \re{metric}, the energy-momentum
tensor has the form
\bea \label{emt}
  T^{\mu}_{\ \nu} = \textrm{diag}( -\rho(t), p(t), p(t), p(t), P(t), P(t),  P(t),  P(t),  P(t),  P(t) ) \ .
\eea

\noindent With \re{metric} and \re{emt}, the Einstein equation \re{eom} reads
\bea
  \label{Hubble} \kappa^2\rho &=& 3 \HH + 18 \H\Hb + 15 \HHb \\
  \label{addot} \Hd &=& - \frac{1}{6} \kappa^2 ( \rho + 3 p ) - \frac{3}{8} \kappa^2 ( \rho - 3 p + 2 P ) + 6 \H\Hb + 10 \HHb \\
  \label{bddot} \kappa^2 ( \rho - 3 p + 2 P ) &=& 8 \Hbd + 24 \H\Hb + 40 \HHb \ ,
\eea

\noindent where $\H\equiv\adot/a$ is the expansion rate 
of the three visible dimensions and $\Hb\equiv\bdot/b$
is the expansion rate of the six extra dimensions.

When the extra dimensions are static, $\Hb=0$,
we recover the usual FRW equations in the visible directions.
In order to have $\Hb=0$, the driving term of $b$ must vanish,
$\rho - 3 p + 2 P = 0$.

\paragraph{The matter content.}

We will consider six kinds of matter. Ordinary four-dimensional
radiation ($\gamma$), ordinary four-dimensional matter, also called
dust ($d$), (consisting of baryons ($b$) and cold dark matter ($cdm$))
and vacuum energy ($\Lambda$) contribute to the energy-momentum
tensor \re{eom} with
\bea 
  \label{gamma} && \rho_{\gamma} = \rho_{\gamma,in} a^{-4} b^{-6} \ , \quad\qquad p_{\gamma} = \frac{1}{3} \rho_{\gamma} \ , \quad\quad \ P_{\gamma} = 0 \\
  \label{baryon} && \rho_{b} = \rho_{b,in} a^{-3} b^{-6} \ , \quad\quad\quad p_{b} = 0 \ ,  \quad\quad P_{b} = 0 \\
  \label{cdm} && \rho_{cdm} = \rho_{cdm,in} a^{-3} b^{-6} \ , \quad p_{cdm} = 0 \ ,  \quad P_{cdm} = 0 \\
  \label{Lambda} && \rho_{\Lambda} = - p_{\Lambda} = - P_{\Lambda} \ ,
\eea

\noindent extra-dimensional dark matter ($edm$) obeys
\be
  \label{edm} \rho_{edm} = \rho_{edm,in} a^{-3} b^{-3} \ , \quad p_{edm} = 0 \ , \quad P_{edm} = - \frac{1}{2} \rho_{edm} \ ,
\ee

\noindent and finally, the string gas ($s$) has
\bea
  \label{rhos} &&\rho_s = M^{-1} \rho_{s,in} a^{-3} b^{-6} \sqrt{ M^2 a^{-2} + b^{-2} + b^2 - 2 } \\
   \label{ps} && p_s = \frac{1}{3} \frac{M^2 a^{-2}}{ M^2 a^{-2} + b^{-2} + b^2 - 2 } \rho_s \\
  \label{Ps} && P_s = \frac{1}{6} \frac{ b^{-2} - b^2 }{M^2 a^{-2}  + b^{-2} + b^2 - 2 } \rho_s \ ,
\eea

\noindent where the subscript $in$ refers to the initial values,
and $M$ is the average initial energy of a string
(due to momentum in the visible directions) in units of the
string length $l_s$. (See
\cite{Patil:2004, Patil:2005a} regarding the
extra-dimensional dark matter, and the appendix for the
derivation of the string gas energy-momentum tensor.)

Baryons and dark matter, both $cdm$ and $edm$, are always
pressureless in the three large dimensions and will be
collectively called matter ($m$),
$\rho_m\equiv\rho_b+\rho_{cdm}+\rho_{edm}$. 
Four-dimensional matter, composed of baryons and cold dark matter,
will be collectively called dust, $\rho_d\equiv\rho_b+\rho_{cdm}$. 
We introduce the dust fraction $f_d\equiv \rho_d/\rho_m$
to measure how much of the matter is four-dimensional.
If all dark matter is extra-dimensional, $f_d=f_b\equiv\rho_b/\rho_m$,
and if all dark matter is four-dimensional, $f_d=1$.
In principle, we could also have a mixture of $cdm$ and $edm$,
$1>f_d>f_b$.

\subsection{Dust and strings} \label{sec:dust}

\paragraph{Destabilising dust.}

We are interested in how the size of the extra dimensions
behaves in the late-time universe, starting from the
radiation dominated era, evolving to being dominated
by dust and finally by vacuum energy.

When the universe is radiation dominated, the
stability condition \mbox{$\rho - 3 p + 2 P = 0$}
is satisfied, and the extra dimensions stay at the
self-dual radius, $b=1$. When dust becomes
important, the driving term $\rho - 3 p + 2 P$
becomes positive and $b$ grows. From the
four-dimensional point of view, the expansion of the
extra dimensions looks like a time-dependent Newton's
constant, with $8\pi G_N = \kappa^2 l_s^6 b^{-6}$.
The change of Newton's constant from the initial
value $G_{N,in}$ during BBN to the value today $G_{N0}$
is constrained to be $G_{N0}/G_{N,in}=0.98^{+0.25}_{-0.18}$
(2$\sigma$ limit using $^4$He and D abundance, and assuming
negligible neutrino chemical potential) \cite{Cyburt:2004},
which translates into $b_0=1.00^{+0.04}_{-0.03}$.
This bound applies only to the final value,
and does not limit $b$ from being large between BBN and today.
Allowing for a non-negligible neutrino chemical potential,
a combined analysis of BBN and the cosmological microwave
background (CMB) leads to $1.38>G_{N0}/G_{N,in}>0.60$,
which translates into $1.09>b_0>0.95$
(2$\sigma$ limit using older values for $^4$He and D
abundance) \cite{Barger:2003, Fields:2004}.
Using information from the CMB makes the limit more
model-dependent: in particular, the radiation degrees
of freedom are assumed to be the same during BBN and
at last scattering (which is in fact not true in
the string gas model). Limits from CMB,
large-scale structure and globular clusters \cite{CMB}
constrain the value of $b$ between last scattering
and today, but are strongly model-dependent.
Even with the limits from BBN, we should take into account that the
string gas provides additional radiation degrees of
freedom, which can either tighten or relax the bounds
(depending on how much the string gas contributes to the
energy density during BBN and whether $b$ is larger or
smaller than unity today).

One caveat concerning the above limits is that
the Newton's constant measured on Earth is not
necessarily the cosmologically relevant quantity.
This is both because the size of the extra dimensions
may behave differently in regions where the visible
dimensions are expanding and in regions which have
broken away from the general expansion \cite{Clifton:2004},
and because the $b$-dependence in the Einstein equation
does not factorise for all forms of matter. The
latter point simply expresses the fact that the division
between energy density and Newton's constant is one of
convention\footnote{For example, $\rho_{edm}\propto b^{-3}$ in
\re{edm}, so we could say that the effective four-dimensional
gravitational coupling of the extra-dimensional
matter goes like $b^{-3}$, in contrast to the $b^{-6}$
behaviour of ordinary matter. Alternatively, we could
say that the gravitational coupling is the standard one,
but the energy density has an additional factor of $b^3$.}
(see \cite{constants} for discussion).

In \cite{Patil:2004, Patil:2005a} it was concluded that
because four-dimensional dark matter destabilises the
extra dimensions, dark matter in SGC has to be
extra-dimensional, i.e. if the pressure in the visible
directions is zero, the pressure in
the extra directions has to be negative to keep $b$ static.
Strings with winding around (but no momentum in) the extra
dimensions were suggested as a dark matter candidate satisfying
$p_{edm}=0$, $P_{edm}=-\frac{1}{2}\rho_{edm}$, as listed
in \re{edm}, so that \mbox{$\rho_{edm} - 3 p_{edm} + 2 P_{edm}=0$}.

However, even if all dark matter was extra-dimensional,
there would still be the four-dimensional baryons (as 
well as vacuum energy, which we
discuss in section \ref{sec:acc}),
with $p_b=P_b=0$, to destabilise
the extra dimensions. Note that while the
extra-dimensional dark matter does not destabilise the
extra dimensions, neither does it help to stabilise them.
Figure \ref{fig:dust1} shows the evolution of a model
with baryons and extra-dimensional dark matter
(no vacuum energy). In this case the dust fraction is just
the baryon fraction, which we take to initially be
$f_d=f_b=0.17$ (since $\rho_b/\rho_{edm}\propto b^{-3}$,
the baryon fraction evolves with time).
The extra dimensions start growing logarithmically
after baryons become dynamically important.
The size of the extra dimensions today (taken to be at 13.7 billion years,
marked with the vertical line) is $b_0=1.3$, far in excess
of the limits quoted above. However, the growth of the extra dimensions
due to baryons can be checked by the same gas of strings which
stabilises the extra dimensions in the early universe.

\FIGURE[b,t]{
\includegraphics[width=\textwidth,height=0.55\textheight]{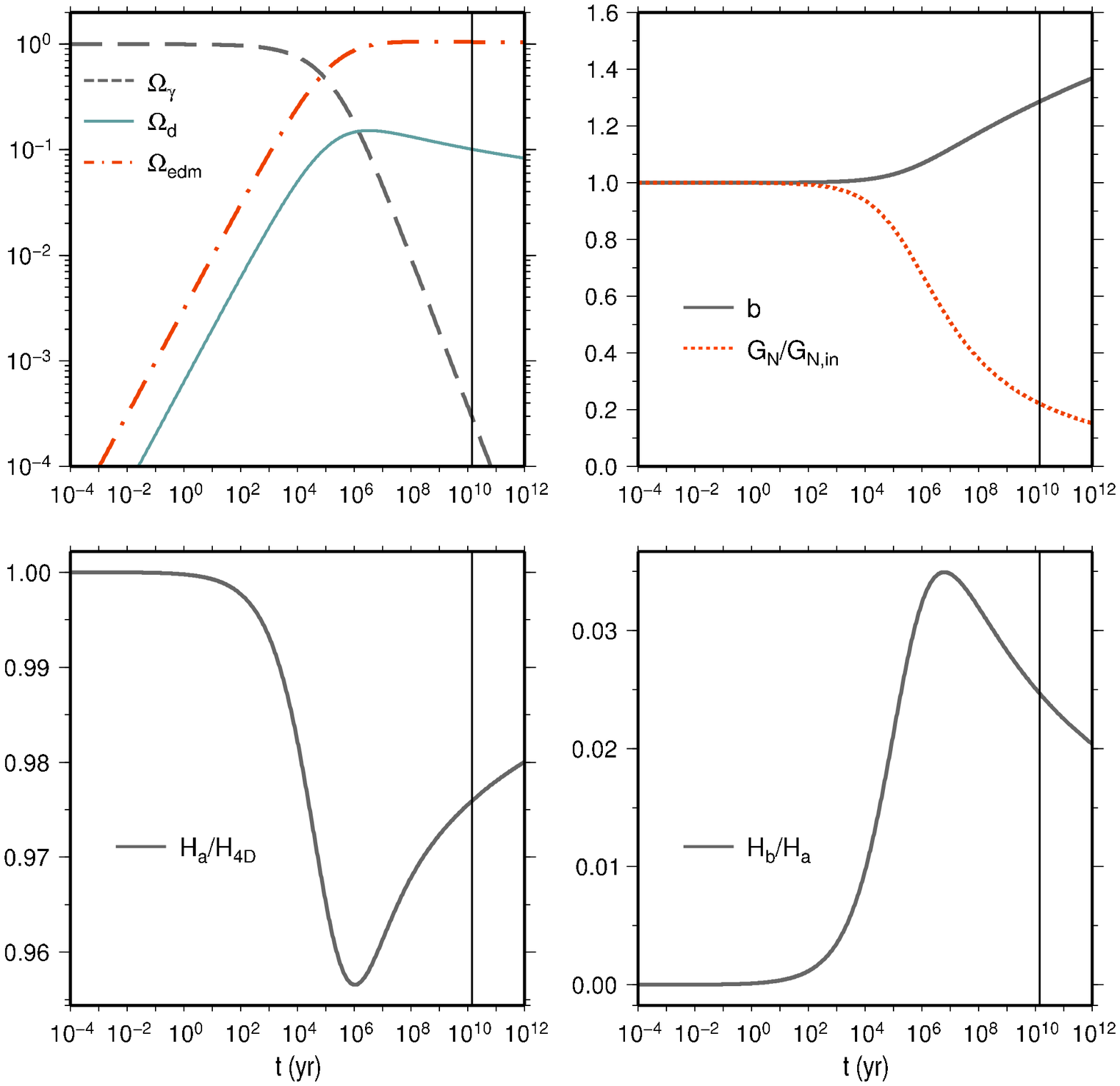}
\caption{Density parameters $\Omega_i \equiv \kappa^2\rho_i/(3 H_a^2)$ (top left),
the size of the extra dimensions and Newton's constant (top right),
expansion rate of the large dimensions ($H_{4D}$ is the
Hubble parameter in the usual four-dimensional case) (bottom left)
and the expansion rate of the extra dimensions (bottom right).
All dark matter is extra-dimensional ($f_d=f_b=0.17$) and there
is no vacuum energy. The vertical line marks the present time $t_0=13.7$ Gyr.}
\label{fig:dust1}
}

\paragraph{Stabilising strings.}

The mechanism for driving the extra dimensions to the self-dual
radius relies on a gas of strings having momentum and winding
around the extra dimensions. The contribution of the string
gas to the driving term of $b$ is, from \re{rhos}--\re{Ps},
\bea \label{sdrive}
  \rho_s - 3 p_s + 2 P_s = \frac{2}{3} \frac{ 2 b^{-2} + b^2 - 3 }{ M^2 a^{-2} + b^{-2} + b^2 - 2 } \rho_s \ .
\eea

During the radiation dominated era the extra dimensions
remain static at \mbox{$b=1$}, and the stabilising strings behave
like four-dimensional radiation\footnote{Their initial density is thus constrained
by the limit on additional radiation degrees of freedom
during BBN: $\Omega_{s,in}<0.20$ with zero neutrino chemical
potential \cite{Cyburt:2004}, $\Omega_{s,in}<0.40$
with a non-zero neutrino chemical potential \cite{Barger:2003}
(both 2$\sigma$ limits).}, as \re{rhos}-\re{Ps}
show,
so their contribution to the driving term is zero.
As the contribution of baryons to the driving term
becomes important and pushes $b$ up, the string
contribution \re{sdrive} becomes negative and tries to
drive $b$ downward (should $b$ dip below 1, the string
contribution will change sign and push $b$ back up).

\FIGURE[b,t]{
\includegraphics[width=\textwidth,height=0.55\textheight]{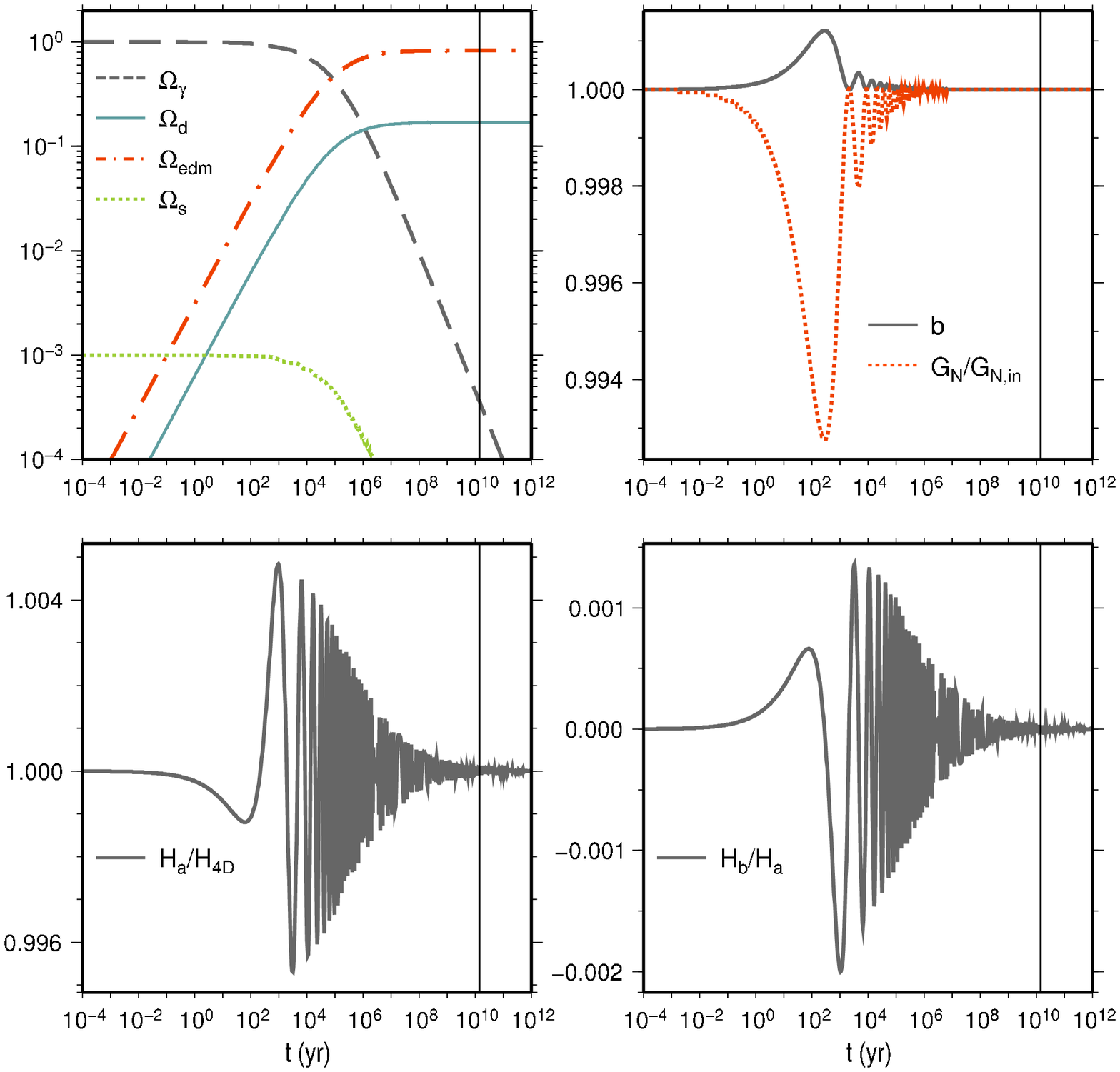}
\caption{The model of Figure \ref{fig:dust1}, with $f_d=0.17$,
now with stabilizing strings added. The initial density of the strings
is $\Omega_{s,in}=10^{-3}$ and $M=588$.}
\label{fig:dust2}
}

Two observations may be made about the stabilisation mechanism.
First, $b=\sqrt{2}$ is a point of no return. If $b$ grows beyond
$\sqrt{2}$, the string contribution to the driving term becomes
positive, so there is nothing to stop further expansion and
the extra dimensions will decompactify.
Second, there is no static solution for $b$. If the strings
drive the extra dimensions back to the self-dual radius,
baryons will destabilise them again. The extra dimensions either
grow indefinitely or oscillate around the self-dual radius, but
they cannot remain fixed at the self-dual radius.

Figure \ref{fig:dust2} shows the same model as before,
with extra-dimensional dark matter and baryons (and no
vacuum energy), but now with stabilising strings added.
As expected, $b$ oscillates near the self-dual radius,
with a decreasing amplitude. Note that the strings
can effectively stabilise the extra dimensions against
the destabilising effect of baryons even when
the contribution of strings to the total energy density
is negligible (of the order $10^{-3}$ or less) throughout.

With regard to the dynamics of the extra dimensions,
there is no difference between four-dimensional baryons
and four-dimensional dark matter. So, the string gas
can cure the destabilising effects of cold dark matter
as well. Figure \ref{fig:dust3} shows the same model as before,
but with extra-dimensional dark matter replaced with cold dark matter.
Though the destabilisation is stronger than in the
case with extra-dimensional dark matter, qualitatively
the behaviour is the same as before.

\FIGURE[b,t]{
\includegraphics[width=\textwidth,height=0.55\textheight]{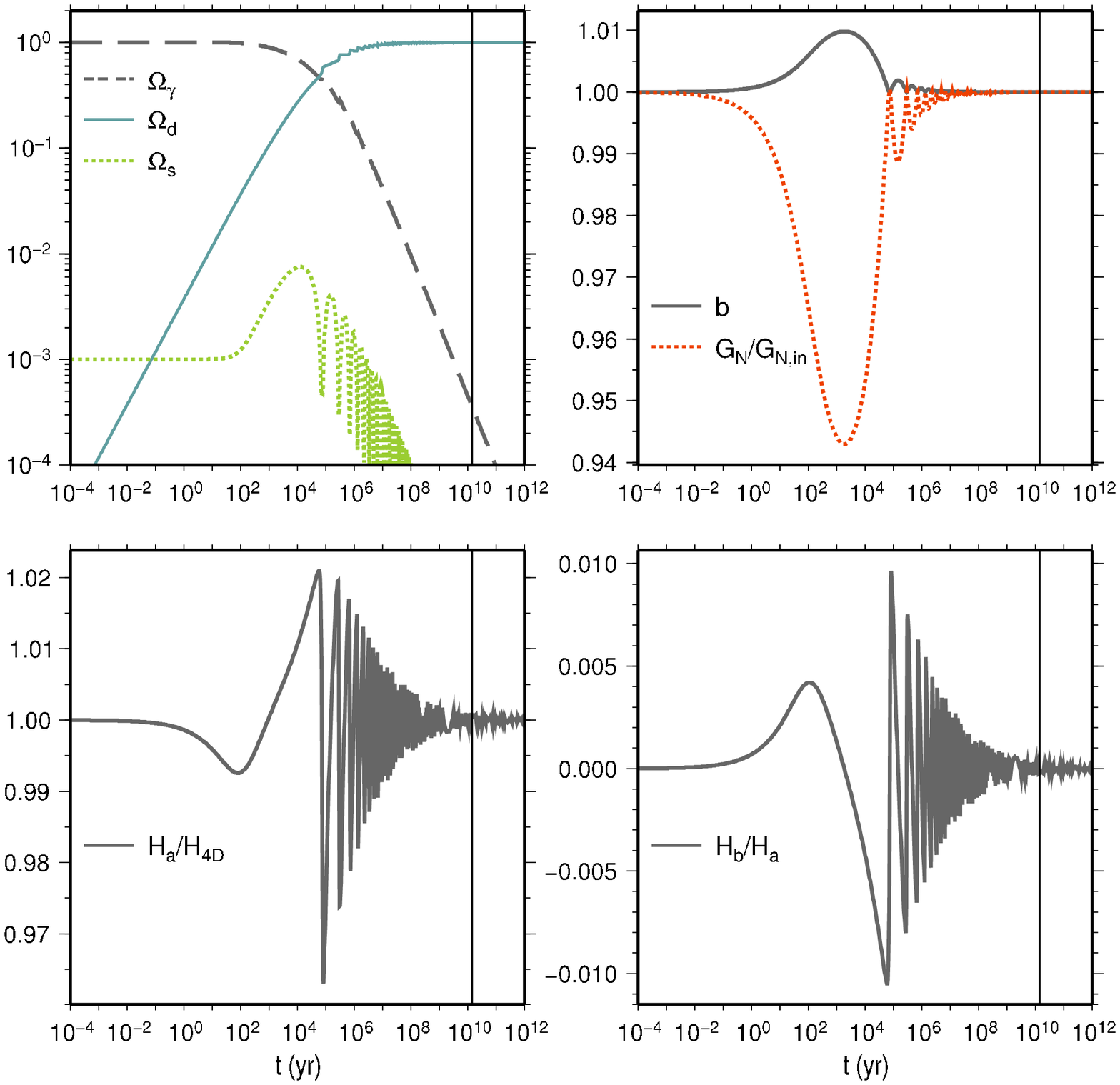}
\caption{The same model as in Figure \ref{fig:dust2}, but now with
cold dark matter instead of extra-dimensional dark matter, $f_d=1$.}
\label{fig:dust3}
}

Requiring that the strings turn the driving term of $b$
negative before the point of no return at $b=\sqrt{2}$
leads to the limit
\bea \label{Mlimit}
  \frac{M^{-1} \rho_{s,in}}{\rho_{m,in}} > \frac{3}{2} f_d \ .
\eea

\noindent This limit is a necessary condition for stabilisation:
unless \re{Mlimit} satisfied, the string gas contribution is too
weak to overcome the baryons, and the extra dimensions will
decompactify. However, it is not a sufficient condition,
because if the driving term becomes negative too close to the point
of no return, there isn't enough time to turn $b$ around.

It may seem paradoxical in \re{Mlimit} that the strength
of the stabilisation increases with decreasing initial string
energy $M$. The reason is that, for a constant energy density,
smaller energy means larger number density. When $b$ is displaced
from the self-dual radius, the energy that the extra-dimensional
momentum and winding modes of a single string
contribute is a fraction of the string scale $l_s^{-1}$,
and does not depend on $M$. So, the stabilising contribution
is proportional to the number density of strings and independent
of their initial energy density.

To summarise, we find that SGC with cold dark matter can produce
a matter-dominated era in agreement with observations, in contrast
to the conclusion of \cite{Patil:2004, Patil:2005a}.
The same strings which stabilise the extra dimensions in the early
universe can effectively stabilise them during the matter dominated era.
Extra-dimensional dark matter then seems like an unnecessary complication.
Having observed that dust poses no problem for late-time
cosmology, we now take a look at vacuum energy, and
discuss the relation between accelerated expansion of
the visible dimensions and destabilisation of the extra dimensions.

\subsection{Destabilisation and acceleration} \label{sec:acc}

\paragraph{Vacuum energy.}

Vacuum energy obeys the equation of state
$p_{\Lambda}=P_{\Lambda}=-\rho_{\Lambda}$,
so its contribution to the driving term is
positive and will decompactify the extra dimensions.
Figure \ref{fig:lambda} shows the evolution of a model with
dust ($f_d=1$) and vacuum energy in the 'concordance'
proportions, and the same string gas as in Figure \ref{fig:dust1}.
(We take the 'concordance model'
values $\Omega_m=0.27$, $\Omega_{\Lambda}=0.73$ at
13.7 billion years \cite{Spergel:2003} and extrapolate
back to the BBN era when we give our initial conditions.)

\FIGURE[b,t]{
\includegraphics[width=\textwidth,height=0.55\textheight]{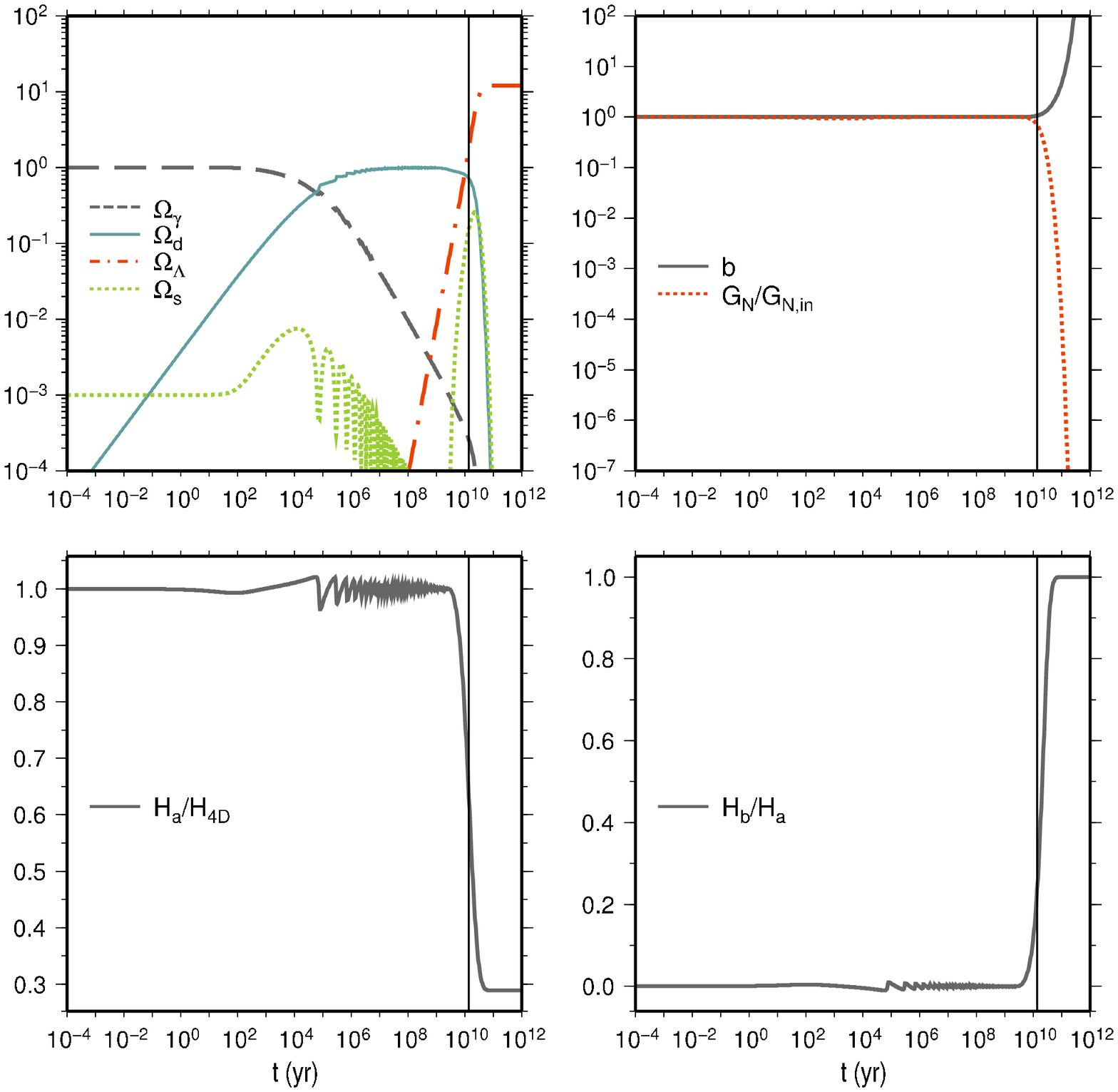}
\caption{A model with 'concordance' proportions of dust ($f_d=1$)
and vacuum energy, and the same stabilising strings as in Figures
\ref{fig:dust2} and \ref{fig:dust3}.}
\label{fig:lambda}
}

The initial small oscillations of $H_a$ and $H_b$ are due
to the interplay of dust and the string gas, and are the
same as seen in Figure \ref{fig:dust3}. When vacuum
energy becomes important, $b$ starts growing exponentially
and the visible and extra dimensions rapidly isotropise
as they undergo exponential inflation, in agreement with
the no-hair conjecture \cite{nohair}.
As in the case of baryons, the growth of $b$
can in principle be used to rule the model out. For
the concordance values, we have $b_0=1.06$ today, inside
the $2\sigma$ limit from BBN (when allowing for a neutrino
chemical potential). Tighter bounds on the
electron neutrino chemical potential, a larger age of
the universe (or perhaps simply redoing the neutrino
chemical potential analysis of \cite{Barger:2003} with the
updated abundances in \cite{Cyburt:2004}) could rule the model out.

In contrast to the gentle push from baryons
which led to logarithmic growth of $b$, the dramatic
destabilisation due to vacuum energy cannot be prevented by the string gas.
In fact, destabilisation of the extra dimensions is a general
feature of acceleration of the visible dimensions.
Assuming that the extra dimensions are static implies
\mbox{$\rho-3p+2P=0$}. Acceleration in the visible dimensions then
requires $\rho+3p<0$, which leads to $\rho+P<0$. So, acceleration
implies that the extra dimensions are dynamical or the
null energy condition is violated or both. This is in
agreement with the observation in \cite{Patil:2005a} that a
period of scalar-field driven inflation would destabilise
the extra dimensions. (For discussion of inflation in SGC, see
\cite{Brandenberger:2003, Kaya:2004, Biswas:2005a, Easson:2005}.)

It might seem that SGC is under stringent constraints due
to the change in $b$ implied by acceleration\footnote{Unless
observations could be fitted without acceleration, see
\cite{Evslin:2005} for an interesting suggestion and also
\cite{Sarkar}.}, and that it would be ruled out by a small
tightening of the bounds -- at least in the simplest setting
where the extra dimensions are toroidal, isotropic, spatially flat
and stabilised by the string gas (changing vacuum energy to
a dark energy model with more parameters could also
allow some breathing room). However, there is a surprising way out:
though the stabilising strings cannot prevent
decompactification in the case of vacuum energy,
they can lead to acceleration without decompactification
in the absence of vacuum energy (though this will turn out to
involve violating the null energy condition).

\FIGURE[h,b,t]{
\includegraphics[width=\textwidth,height=0.55\textheight]{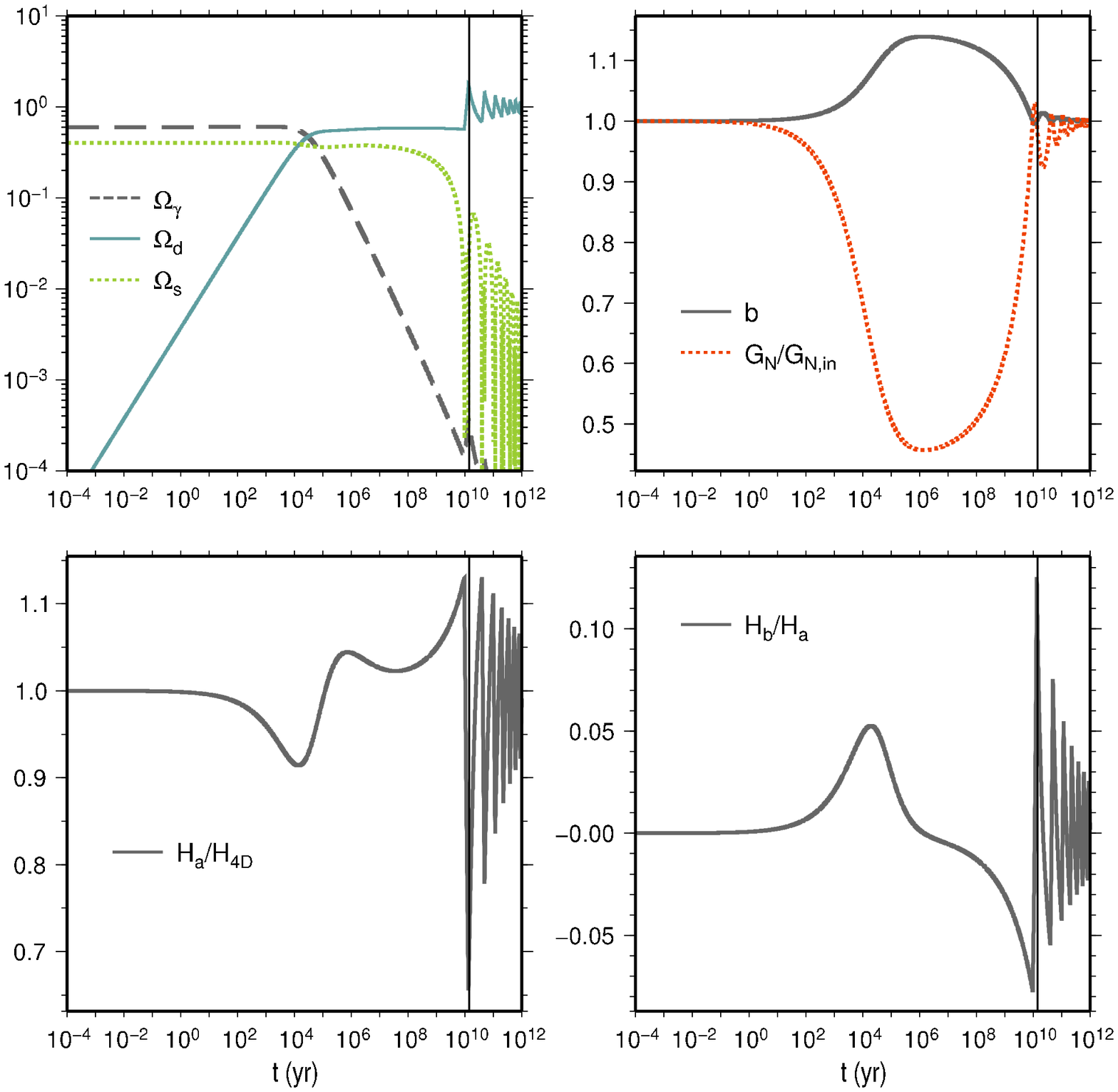}
\caption{An accelerating model with dust ($f_d=1$) and strings
($\Omega_{s,in}=0.40$, $M=158\,730$).}
\label{fig:acc1}
}

\paragraph{Dust.}

Acceleration leads to destabilisation of the extra
dimensions, so we can ask whether it is possible to obtain
acceleration as a result of destabilising the extra
dimensions, rather than vice versa. Let us assume that
the vacuum energy is zero. As dust
becomes dynamically important, it will push the extra
dimensions to expand. The stabilising strings will
pull the extra dimensions back and will eventually
turn $b$ around (if their number density is large enough).
The acceleration equation \re{addot} shows that
a negative driving term for $b$ contributes positively to
the acceleration of $a$, suggesting that a collapsing extra
dimension could lead to acceleration. There is also a negative
contribution to the acceleration from the terms involving $H_b$
(for $|H_b|<3/5 H_a$), so it is not immediately clear what will happen.

\FIGURE[h,t]{
\includegraphics[width=\textwidth,height=0.55\textheight]{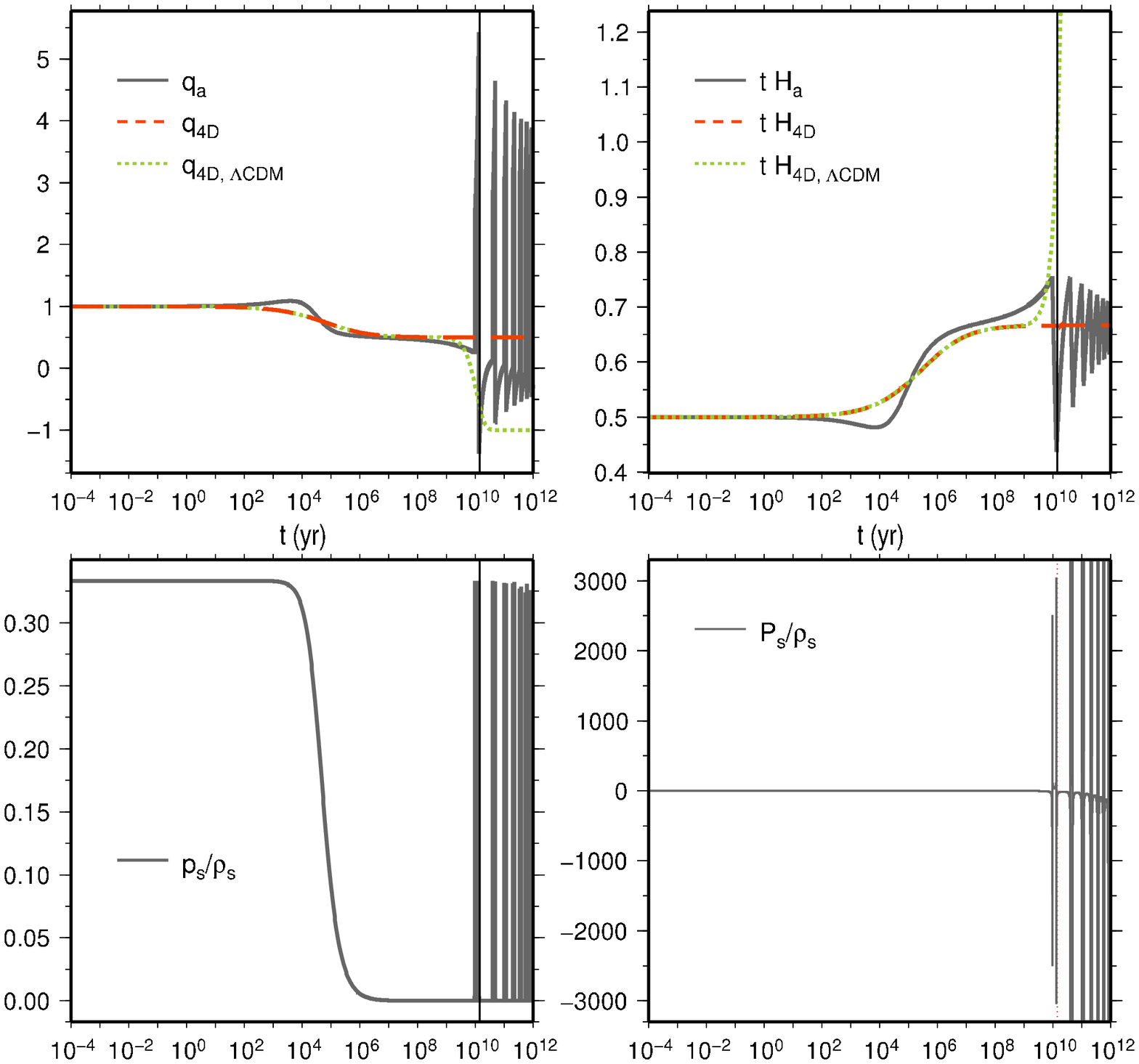}
\caption{The deceleration parameter (top left), the expansion
rate of the visible dimensions with regard to age (top right),
pressure in the visible dimensions (bottom left),
pressure in the extra dimensions (bottom right).
The model is the same as in Figure \ref{fig:acc1}.}
\label{fig:acc2}
}

Figure \ref{fig:acc1} shows the evolution of a model with
baryons, cold dark matter ($f_d=1$) and stabilising strings.
As in Figures \ref{fig:dust2} and \ref{fig:dust3}, the
strings can have a major impact on the dynamics even when
their contribution to the total energy density is negligible
($\rho_s/\rho<10^{-2}$ today).
First the extra dimensions open up and the expansion
rate of the visible dimensions slows down (relative to the
4D case). When the extra dimensions turn around, the expansion
rate of the visible dimensions speeds up as $b$ collapses,
and reaches a maximum at the minimum value of $H_b$. As $b$
dips below 1, the strings rapidly bounce it back,
and $b$ starts oscillating around the self-dual radius
as in Figures \ref{fig:dust2} and \ref{fig:dust3}.
The amplitude of the oscillations decreases rapidly, and today we have
$b_0=1.002$, within the observational limits
(even when also taking the contribution of the string
energy density into account).
However, now the oscillations also involve going back and
forth between acceleration and deceleration. The deceleration
parameter $q=-\addot/a/H_a^2$ is shown in the
top left panel of Figure \ref{fig:acc2}.

Not only can the string gas stabilise the extra
dimensions in the matter-dominated era, but it can also
lead to late-time acceleration in the process. This seems
particularly noteworthy given that we have added no new
ingredients to SGC, simply taken account of the fact
that baryons and cold dark matter are four-dimensional.
However, the rapidly oscillating behaviour shown in
Figures \ref{fig:acc1} and \ref{fig:acc2} is quite
different from the smooth change from deceleration to
acceleration in the $\Lambda$CDM model, and we will now
discuss the comparison to observations.

\section{Discussion}

\subsection{Phenomenology}

In order to compare the model to observations, 
one has to solve the equations \re{Hubble}--\re{bddot}
with the matter content given by \re{gamma}--\re{Ps}
numerically for each value of
$\Omega_{s,in}$, $M$ and $f_d$ to obtain $a(t)$ to compare
with observations (and $b(t)$ to check it is not excluded
by observations). Comparison to SNIa data would be relatively
straightforward, while comparison to CMB and large-scale
structure would require extending the equations beyond
the homogeneous level to perturbation theory \cite{Watson:2003uw,Watson:2004vs, Battefeld:2005}.
We leave a thorough study of the $(\Omega_{s,in}, M, f_d)$-parameter
space and detailed comparison to observations for later work,
and only make some mostly qualitative comments here.

Since the extra dimensions start opening up when
matter becomes dynamically important, the acceleration
naturally starts only after the matter-radiation equality.
(Also, it is more difficult for oscillations to reach into the
accelerating regime during the radiation-dominated era because
$q$ is larger.) This is reminiscent of tracker models \cite{tracker},
and the strings do go from behaving like radiation to behaving
like matter as the universe goes from radiation domination to matter
domination, if the growth of $b$ is slow (as one can see from the
plot of $p_s/\rho_s$ in the bottom left panel of
Figure \ref{fig:acc2}, or by expanding \re{rhos} around $b=1$).
With regard to the coincidence problem, the model has the same
shortcoming as tracker fields: the preferred time is shortly
after the matter-radiation equality at $t_{eq}\sim 10^5$ years,
but according to observations the acceleration starts much later,
at around 10 billion years $\sim 10^5 \, t_{eq}$.

The acceleration starts as the collapse of the extra dimensions
ends with $b$ starting to oscillate. So, the later $b$ turns
around from expansion from collapse, the later the acceleration
starts. The latest possible turnaround is achieved for models
where $b$ is at the limit of expanding forever instead of 
turning around. A large peak value of $b$ also seems to help
by making the end of the collapsing period violent enough that
the resulting oscillations reach deep into the accelerating
domain, as required by observations ($q_0\leq-0.3$ \cite{Riess:2004}).
For the accelerating model
shown in Figure \ref{fig:acc1} with $M=158\,730$, the extra
dimensions would decompactify for the value $M=160\,000$,
and for smaller values the acceleration would start earlier
(for example, for $M=50\,000$ it starts between $10^5$ and $10^6$ years).
So, while it is natural for the acceleration to be
in the late era of cosmology, starting as late as
$\sim 10^5\, t_{eq}$ requires tuning.

Another possibility could be that the oscillations
have indeed started early, and that we are not seeing the
first stage of acceleration. However, in the stabilised
models we have looked at, the quantity $t H_a$ stays near the 3D
matter-dominated value 2/3 already after a few oscillations,
as shown in the top right panel of Figure \ref{fig:acc2}. In
other words, the periods of acceleration and periods of extra
deceleration cancel each
other out. The value today $t_0 H_{a0}$ can be written as
$1.3 h \times t_0 / (13 \textrm{ billion years})$,
where $H_{a0}=h$ 100 km/s/Mpc, which for the 'concordance'
values $h=0.71$, $t_0=13.7$ billion years gives 0.99.
In Figure \ref{fig:acc2}, the maximum value of $t H_a$ is only 0.75
at the first oscillation, going down thereafter. Further, the
minima of $t H_a$ and $q$ coincide, so getting a large enough
$t H_a$ simultaneously with a negative enough $q$ to match
observations is increasingly difficult for later oscillations. 
Even for the first transition to acceleration it looks
desirable to shift the age of the universe to optimise
the fit of $q$ and $t H_a,$\footnote{Though see \cite{Sarkar}
regarding the observational value of the Hubble parameter.}
though we emphasise that we have not done a comprehensive
search of the parameter space to find the best-fitting model.
Note that most parametrisations used to fit the SNIa data
would not see rapid oscillations, in which case we should rather
consider the average over oscillations (the scatter of the
unbinned data points is large, so it is not clear how well
one can detect small-frequency oscillations with the current data).
If the oscillations are not too rapid to see individually in
the data, then the sharp transition to acceleration, whether
it is the first or a subsequent one, is a distinctive signature.
A sharp transition is allowed, but not required, by the current
data~\cite{trans,alam,trans2,hann}. Note that constraints
derived for a rapid transition (see e.g.~\cite{padm}) often depend
on the chosen parametrisation for the
equation of state~\cite{trans2,hann}.

The values of $M$ we have used correspond to very high energies:
the strings used in Figures \ref{fig:acc1} and \ref{fig:acc2}
have an average momentum
of $M\sim 10^5 l_s^{-1}$ at the BBN era. At first sight, this may seem
unnatural, since for point particles in thermal equilibrium
the number density of high-energy states goes down exponentially.
However, for a gas of strings at high energies in a compact space, the existence
of winding modes makes the energy concentrate in a small number
of highly energetic strings \cite{Deo:1989a, Deo:1989b, Deo:1992},
as discussed in the appendix. It is amusing that
instead of having to input an unnaturally low energy scale
for the acceleration to start late enough, as in most
dark energy models\footnote{See \cite{Biswas:2005b} for an
interesting idea for avoiding the introduction of a low energy scale.},
we need an extremely high energy scale, which arises naturally
in string thermodynamics in compact spaces.

Another parameter which seems large is the initial energy density
of strings, which we have pushed to the limit allowed by BBN
bounds (to make the effect of the strings clearer).
From the limited studies of the parameter space we have done,
it seems that the minimum value of $q$ depends on $\Omega_{s,in}$
(unlike the strength of the stabilisation, which depends on the
number density). For $f_d=1$, at least $\Omega_{s,in}=0.10$
still results in acceleration.

It is noteworthy that the deceleration parameter can dip
below the de Sitter value $-1$. Such rapid acceleration is
usually associated with violation of the null energy condition,
i.e. equations of state more negative than $-1$.
Since the departure from matter-dominated 4D behaviour is due
to both the extra dimensions and the string gas, the
apparent dark energy does not obey a simple equation of state.
However, it is still true that the violation of the null energy
condition by the strings is an essential ingredient of the
acceleration (it allows the strings to have a major impact on
the dynamics even when their energy density is negligible).
The bottom right panel in Figure \ref{fig:acc2} shows
$P_s/\rho_s$, which reaches values of over 100 and below -720
in the period until today. In \cite{Patil:2004, Patil:2005a}
it was argued that the dominant energy condition is not
violated because the strings have a large momentum in the
visible directions (compared to $\sqrt{b^{-2}+b^2-2}\, l_s$).
The violation of the dominant and null energy conditions
shown in Figure \ref{fig:acc2} is indeed related to the
momentum in the visible directions becoming small compared
to $\sqrt{b^{-2}+b^2-2}\, l_s$, as can be seen by expanding
\re{Ps} around $b=1$.

An interesting feature related to the violation of the null
energy condition is that ordinary matter dominates the energy
density even when the universe is accelerating, so we could
in principle have $\Omega_{tot}\approx\Omega_m=0.2\ldots0.3$
today (this requires $H_b/H_a=-0.13\ldots-0.15$, and is not
realised in the model shown in Figures \ref{fig:acc1}
and \ref{fig:acc2}).
However, this does not imply spatial curvature, since the
correspondence between spatial flatness and critical density
is broken by the extra-dimensional terms in the Hubble law \re{Hubble}:
$\Omega_m\equiv\kappa^2\rho_m/(3 H_a^2)\neq\rho_m/\rho$.

\subsection{Conclusion}

We have studied the stabilisation of the extra dimensions in
late-time string gas cosmology (SGC). In addition to
the destabilisation by four-dimensional dark matter noted in
\cite{Patil:2004, Patil:2005a}, baryons will drive the
extra dimensions to expand. The effect of both baryons and
four-dimensional dark matter can be checked by the string gas
which has stabilised the extra dimensions in the early universe.
However, vacuum energy, or any other dark energy
candidate satisfying the null energy condition, will
rapidly decompactify the extra dimensions.

We find that SGC has a built-in mechanism for producing late-time
acceleration which doesn't decompactify the extra dimensions or
require vacuum energy.
The interplay between matter pushing the extra dimensions
to expand and strings reining them in leads to oscillations
around the self-dual radius. This can involve
oscillations between deceleration and acceleration,
depending on the number density and energy density
of the string gas. The violation of the null energy
condition by the strings makes it possible
to have acceleration even when the energy density
of the universe is dominated by ordinary matter, with
$\Omega_{tot}\approx\Omega_m<1$, without contradicting
spatial flatness.

Showing that a matter-dominated period followed by
accelerated expansion without decompactification
is possible may be seen as a step towards developing
SGC into a realistic model of the universe at all eras.
However, it is not clear whether the late-time acceleration produced
by this mechanism can be in agreement with observations, and
we leave a detailed study of the parameter space of the model
and comparison to observations for future work. 

\ack

We thank Robert Brandenberger and Subodh Patil for correspondence.
FF has been supported by the U.S. Department of Energy and NASA at Case.
SR has been supported by PPARC grant PPA/G/O/2002/00479.
Part of this work was done using the Altix 3700 COSMOS supercomputer
from the UK Computational Cosmology Consortium funded by PPARC, HEFCE and
SGI/Intel.

\appendix


\section{Appendix}

In this appendix we derive the energy-momentum tensor of the string
gas, \re{rhos}-\re{Ps}. Following \cite{Patil:2004, Patil:2005a},
we consider a gas of strings with winding and momentum modes in
the extra dimensions, and only momentum modes in the visible
dimensions (the winding modes in the visible dimensions are
assumed to have annihilated). Given the metric \re{metric},
and assuming that all dimensions are compactified on tori,
the energy of a string state is given by\footnote{In the
present context, this quantity is properly called the energy
rather than the mass, since we are not looking at the states from
the four-dimensional Kaluza-Klein point of view.}
\bea \label{energy}
  l_s^2 E^2 = a^{-2} \sum_{i=1}^{3} N_i^2 + \sum_{j=1}^{6} \left( b^{-1} N_j + b W_j \right)^2 + 4 (N-1) \ ,
\eea

\noindent where $l_s\equiv\sqrt{\alpha'}$,
the integers $N_i$ are the momentum quantum numbers of
the visible directions, $N_j, W_j$ are the momentum and
winding quantum numbers, respectively, of the extra
directions, the non-negative integer $N$ is the oscillator level,
and the quantum numbers are subject to the level matching
constraint $N+\sum_{j=1}^{6} N_j W_j\ge0$.
The normalisation is such that $a=b=1$ corresponds to
dimensions at the self-dual radius. In contrast to
\cite{Patil:2004, Patil:2005a}, all momenta are quantised,
since we take all dimensions to be compact (in the late universe
where the visible dimensions
are large and there are no modes winding around them,
this doesn't make any difference).
The validity of \re{energy} requires that $a$ and $b$
change slowly compared to the string scale $l_s$ (see the appendix
of \cite{Patil:2004} for details).

In \cite{Patil:2004, Patil:2005a}, only states with vanishing
contribution to the energy from the oscillator modes and
the extra-dimensional momentum and winding modes at the
self-dual radius (``massless states'') were taken into account. (It
was noted in \cite{Berndsen:2005} that in type II superstring
theory such states are removed by the GSO projection, but
they are present in heterotic string theory.)
Those states which do not remain at zero energy
perturbatively near the self-dual radius were then discarded.
We follow slightly different reasoning. Some of the states with
vanishing energy at the self-dual radius (for example, ones with
$\alpha' E^2 = 3 b^{-2} + b^2 - 4$) become tachyonic away from the
self-dual radius. With the momenta in the visible dimensions
quantised, some states even become tachyonic when the extra
dimensions are stabilised at the self-dual radius and the visible
dimensions expand (for example, ones with
$\alpha' E^2 = 2 a^{-2} + b^{-2} + b^{2} - 4$).
(Some other issues regarding the case when all momenta are
quantised are discussed in \cite{Danos:2004}.)
We keep only the states for which the contribution
to the energy from the oscillator modes and the
extra-dimensional momentum and winding modes
at the self-dual radius vanishes, and which are never tachyonic.
(The requirement of not being tachyonic also removes
the states which have zero energy at half-integer fractions and half-integer
multiples of the self-dual radius, discussed in \cite{Patil:2004}.)
We are left with four sets of quantum numbers:
(1) $N=1, N_j=W_j=0$ for all $j$,
(2) $N=1, N_j=-W_j=\pm1$ for one $j$ and zero for others,
(3) $N=0, N_j=W_j=\pm1$ for one $j$ and zero for others
and (4) $N=0, N_{j_1}=\pm1, W_{j_2}=\pm1$ for two values of
$j_1$ and $j_2$ and zero for others, such that $\sum_{j=1}^{6} N_j W_j=0$.
Excluding tachyonic states leaves almost the same set of states
as requiring zero energy from extra-dimensional and oscillator
modes perturbatively near the self-dual radius. 
(In \cite{Patil:2005a}, the subset of the last category of states
where $j_1\neq j_2$ for all entries were also discarded, since they do
not have zero energy when the perturbations of the extra dimensions
are anisotropic. However, with isotropic extra dimensions, there
is no reason to discard them.) The energy density of a gas of
strings in states with these quantum numbers is
\bea \label{edensity}
  \rho &=& \sum_{\textrm{states}} n_{\textrm{state}}\, E_{\textrm{state}} \el
  &=& \sum_{N^{(1)}_1,N^{(1)}_2,N^{(1)}_3} n_{(1), N^{(1)}_1 N^{(1)}_2 N^{(1)}_3} \sqrt{\sum_{i=1}^3 N_i^{(1)2}} a^{-1} \, l_s^{-1}  \el
  && + 12 \sum_{\{N^{(2)}_i\}} n_{(2), \{N_i^{(2)}\}} \sqrt{\sum_{i=1}^3 N_i^{(2)2} a^{-2} + b^{-2} + b^2 - 2 } \, l_s^{-1} \el
  && + 12 \sum_{\{N^{(3)}_i\}} n_{(3), \{N_i^{(3)}\}} \sqrt{\sum_{i=1}^3 N_i^{(3)2} a^{-2} + b^{-2} + b^2 - 2 } \, l_s^{-1} \el
  && + 300 \sum_{\{N^{(4)}_i\}} n_{(4), \{N_i^{(4)}\}} \sqrt{\sum_{i=1}^3 N_i^{(4)2} a^{-2} + 2 b^{-2} + 2 b^2 - 4 } \, l_s^{-1} \ ,
\eea

\noindent where $n$ is the number density, $N_i^{(q)}$ is the
momentum number in the direction $i=1,2,3$ for a state which
has oscillator and extra-dimensional momentum and winding quantum
numbers identified by $q=1,2,3,4$
(corresponding to the four possibilities listed above),
$n_{q, \{N_i^{(q)}\}}$ is the number density of that state and
the coefficients 12 and 300 are the multiplicities of the states.

In thermal equilibrium, the number density $n_{q, \{N_i^{(q)}\}}$
is given by the occupation number of the state divided by the
volume of the manifold, $l_s^9 a^3 b^6$. We ignore
possible time-dependence of the occupation numbers.
The first sum in \re{edensity} then behaves like four-dimensional
radiation, with all terms proportional to $a^{-1}$. It
brings nothing new to the analysis compared to ordinary
radiation, so we neglect it. The other terms, which include
contributions from both visible and extra dimensions, are more
cumbersome. The sum does not reduce to a single term, but the
various terms are qualitatively the same, so we
replace the sum with a single representative term.
The energy density and the pressures then read
\bea
  \label{rho} \rho &=& \frac{1}{a^3 b^6} n_{s,in} l_s^{-1} \sqrt{ M^2 a^{-2} + b^{-2} + b^2 - 2 } \\
  \label{p} p &=& \frac{1}{3} \frac{1}{a^3 b^6} n_{s,in} l_s^{-1} \frac{M^2 a^{-2}}{\sqrt{ M^2 a^{-2} + b^{-2} + b^2 - 2 }} \\
  \label{P} P &=& \frac{1}{6} \frac{1}{a^3 b^6} n_{s,in} l_s^{-1} \frac{ b^{-2} - b^2}{\sqrt{ M^2 a^{-2}  + b^{-2} + b^2 - 2 }} \ ,
\eea

\noindent where $M$ is the initial average energy of a string
in units of $l_s^{-1}$ and $n_{s,in}$ is the initial number
density, both at the time when $a=b=1$. Identifying
$n_{s,in}=\rho_{s,in}/(M l_s^{-1})$, we have \re{rhos}-\re{Ps}.
The absence of winding modes around the visible dimensions
allows us to rescale $a$ and $M$ so that $a=1$ corresponds to any
convenient era. We have chosen to set $a=1$ at the BBN, when
$\rho_m/\rho_{\gamma}=10^{-6}$ (and \mbox{$\rho_{\gamma}\sim$ (MeV)$^4$}).

Instead of taking a single term, we could have averaged
over the quantum numbers with the correct number density.
The number density
of strings in thermal equilibrium in a toroidal compact space 
where three dimensions grow large has been calculated in
\cite{Deo:1992} (see also \cite{Deo:1989a, Deo:1989b}).
For high energies, the number density falls like (energy)$^{-1}$,
so the relative contribution of high-energy modes to the
energy density is almost independent of their energy
(this conclusion depends on all dimensions, including the large
ones, being compact and admitting one-cycles). The energy is
concentrated in a small number of highly energetic strings,
which is the qualitative picture needed for strings to cause
late-time acceleration. However, the results cannot be applied
directly to the present case, since we have assumed that
the modes winding around the large dimensions have annihilated,
unlike in \cite{Deo:1992}. The number density also depends on
the detailed behaviour of the early universe, particularly on
how the strings are produced and how they thermalise. 
It may be that the temperature at which the
strings are produced is too low for the distributions derived 
in~\cite{Deo:1989a, Deo:1989b,Deo:1992} 
to be relevant. It is also possible that the
strings have never been in thermal equilibrium, and that their number
density is not determined by thermodynamical arguments.
For discussion of thermodynamics in string gas cosmology, see
\cite{Kripfganz:1988, Brandenberger:1989, Sakellariadou:1995, Bassett:2003, Easther:2003, Easther:2004, Danos:2004}.

\end{document}